\documentclass[aoas]{imsart}

\RequirePackage{amsthm,amsmath,amsfonts,amssymb}
\RequirePackage[authoryear]{natbib}
\RequirePackage[colorlinks,citecolor=blue,urlcolor=blue]{hyperref}
\RequirePackage{graphicx, multirow}

\startlocaldefs

\theoremstyle{remark}

\endlocaldefs

\begin{document}

\begin{frontmatter}
\title{A robust Bayesian meta-analysis for Estimating the Hubble Constant  via Time Delay Cosmography}
\runtitle{Robust Meta-Analysis for Hubble Constant Estimation}

\begin{aug}
\author[A]{\fnms{Hyungsuk} \snm{Tak}\ead[label=e1]{tak@psu.edu}}
\and
\author[B]{\fnms{Xuheng} \snm{Ding}\ead[label=e2]{dingxh@whu.edu.cn}}
\address[A]{Department of Statistics, Department of Astronomy \& Astrophysics, Institute for Computational and Data Sciences, The Pennsylvania State University,
\printead{e1}}

\address[B]{Kavli Institute for the Physics and Mathematics of the Universe,
University of Tokyo,
\printead{e2}}

\end{aug}



\begin{abstract} 
We propose a Bayesian meta-analysis to infer the current expansion rate of the Universe, called the Hubble constant ($H_0$), via  time delay cosmography. Inputs of the  meta-analysis are estimates of  two properties for each pair of gravitationally lensed images;  time delay  and  Fermat potential difference estimates with their standard errors. A  meta-analysis  can be appealing in practice because obtaining each estimate  from even a single lens system involves substantial human efforts, and thus estimates are often separately  obtained and  published. \textcolor{black}{Moreover,  numerous estimates are expected to be available once the Rubin Observatory starts monitoring thousands of strong gravitational lens systems.} This work focuses on combining these estimates from independent studies to infer   $H_0$ in a robust  manner. \textcolor{black}{The robustness is crucial because currently up to eight  lens systems are used to infer $H_0$, and thus any biased input can severely affect the resulting $H_0$ estimate.}  
For this purpose, we adopt Student's $t$ error for the input \textcolor{black}{estimates}. 
\textcolor{black}{We investigate properties of the  resulting $H_0$ estimate via two simulation studies with realistic imaging data. It turns out that the meta-analysis can infer $H_0$ with sub-percent bias and about 1\% level of coefficient of variation, even when 30\% of inputs are  manipulated to be outliers. We also apply the  meta-analysis to three gravitationally lensed systems \textcolor{black}{to obtain an $H_0$ estimate and compare it with existing estimates.} An R package \texttt{h0} is publicly available for  fitting the  proposed meta-analysis.} 
\end{abstract}


\end{frontmatter}

\section{Introduction}

Estimates of the Hubble constant $H_0$ under the standard cosmological model, that is, the flat  $\Lambda$ Cold Dark Matter  \textcolor{black}{($\Lambda$CDM)} model,  have been inconsistent, causing a tension between measurements of $H_0$ from  early and  late Universe \citep{verde2019tension, 2021CQGra..38o3001D, 2021A&ARv..29....9S, 2022JHEAp..34...49A}. For example, the most recent probe on the early Universe infers $H_0$ by $67.4 \pm 0.5$ km s$^{-1}$ Mpc$^{-1}$ via  the cosmic microwave background experiment   \citep{planck2020cosmo}. The unit of the Hubble constant is kilometer/second/megaparsec, typically denoted by km s$^{-1}$ Mpc$^{-1}$. It means that if  \textcolor{black}{an object were} 1 megaparsec (about 3.26 million light years) away from the Earth, then \textcolor{black}{it would appear} to move away \textcolor{black}{from us} by 67.4 kilometer per second due to the expansion of the Universe. We omit the unit of $H_0$ hereafter. On the other hand, the \textcolor{black}{most recent} $H_0$ estimate based on a locally calibrated comic distance ladder  (that is, from the late Universe) is $73.04 \pm 1.04$    \citep{2022ApJ...934L...7R}, which is claimed to be 5 standard deviations away from the early Universe study.  These inconsistent estimates between the early and late Universe measurements have raised a question about the validity of the underlying standard cosmological model and brought up a possibility of new physics.  To check whether the tension  is due to  \textcolor{black}{unknown} systematic error of measurements, astronomers have improved data quality and inferential accuracy of existing methods, for example, \cite{riess2019h0},  \cite{2021ApJ...908L...6R}, and \cite{2022ApJ...934L...7R}. \textcolor{black}{Various methods have also been developed under  different physical processes to better understand the  systematic error and test the standard cosmological model.}

\begin{figure}[b!]
\includegraphics[scale = 0.148]{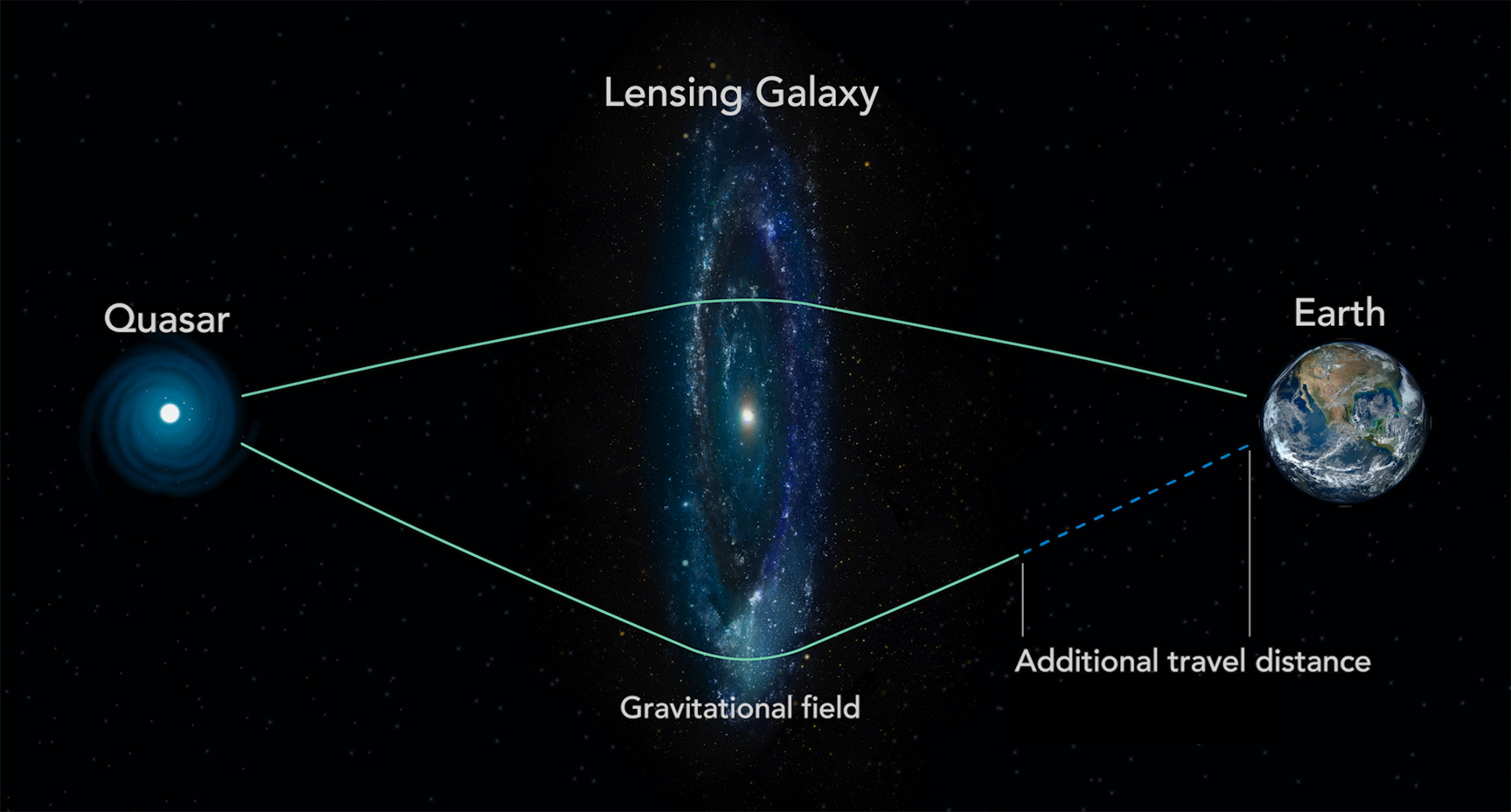}~~~\includegraphics[scale = 0.7]{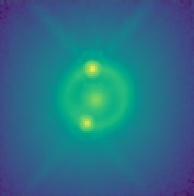}
\caption{Strong gravitational lensing is illustrated in the left panel. When a galaxy is intervening between a quasar and the Earth, light photons emitted by the quasar take  different routes to the Earth because the strong gravitational field of the intervening galaxy bends the trajectories of photons (Image Credit: Michael Fleck). In this case we see multiple images of the same quasar in the sky. The right panel shows simulated imaging data of a doubly-lensed system obtained by  a Python package \texttt{lenstronomy} \citep{lenstronomy, Birrer2021}; the faint image at the center is the intervening lensing galaxy and the two brightest images around the center are the two lensed images of the same quasar.}
\label{fig1lensing}
\end{figure}

As a completely independent way to infer the Hubble constant, time delay cosmography adopts strong gravitational lensing effects of quasars \citep{ref1964h0, Linder2011, treu2016time, suyu2017holicow, 2022arXiv221010833B, 2022A&ARv..30....8T}. When a galaxy is geometrically aligned between a quasar and the Earth, the strong gravitational field of the intervening galaxy bends the trajectories of light photons emitted from the quasar, as illustrated in the left panel of Figure~\ref{fig1lensing}. The deflected light rays travel toward the Earth. As a result  we see multiple (mostly two or four) images of the same quasar in the sky. The right panel of Figure~\ref{fig1lensing} shows  an image of a doubly-lensed strong lens system, simulated by a Python package \texttt{lenstronomy} \citep{lenstronomy, Birrer2021},  as an example of what we expect to see in the sky via a telescope. The faint image at the center is the intervening lensing galaxy, and the two bright images located to the south and north from the center are the two lensed images of the same quasar. We call it a strong gravitational lensing effect \citep{schneider2006, 2010ARA&A..48...87T}. The travel times of  light photons \textcolor{black}{for each lensed image} can be different depending on the paths they take because the length of each path may differ and  photons may pass through different gravitational potential of the intervening  galaxy. We call such differences between their travel times \emph{time delays}.

For each pair of lensed images (one \textcolor{black}{unique} pair for a doubly-lensed system, and 
three \textcolor{black}{unique}  pairs for a quadruply-lensed system), time delay cosmography  models the additional travel distance of the longer route by a physical equation \textcolor{black}{that interweaves cosmological parameters and measurable quantities. Two of the measurable quantities are} time delay and Fermat potential difference, and these \textcolor{black}{can be} separately estimated from two different types of data. Time delay estimates are obtained from several  time series data of brightness of multiply-lensed quasar images  \citep{tewes2013b, eul2013, liao2015tdc, tak2017bayesian, courbin2018des, bonvin2016he, bonvin2018pg,  bonvin2019wfi, millon2020six, 2023ApJ...950...37M}. On the other hand, Fermat potential differences are estimated from high-resolution imaging data of each lens system, \textcolor{black}{such as the one in the right panel of Figure~\ref{fig1lensing}} \citep{lenstronomy,  2020MNRAS.498.1440R,  wong2020,  birrer2020tdcosmo,  shajib2020, tdlmc2020result, 2021MNRAS.504.5621D, 2022arXiv220211101S, 2023A&A...672A...2E, 2023MNRAS.518.1260S}.

These two types of data are independently obtained, \textcolor{black}{and thus} the inferences on  time delays and differences in Fermat potential  \textcolor{black}{can be}   performed separately without knowing each other. \textcolor{black}{This aspect has enabled astronomers to conduct a so-called  blind analysis to estimate each component independently and combine these to infer $H_0$ in the end \citep{2013ApJ...766...70S, suyu2017holicow}}.  Also, it is well known that estimating time delays and Fermat potential differences from even a single lens system  requires substantial human efforts. For instance, each estimation procedure goes through data pre-processing, visual inspection for identifying anomalies, modeling the data with physical processes, model fitting and checking, and interpretation \citep{shajib2019is, 2023MNRAS.522.1323L}. Details of each step \textcolor{black}{are} different according to the types of the data (time series or imaging data). Consequently, these estimates are often published separately in the literature. For instance,  time delay estimates \citep{millon2020six} and Fermat potential difference estimates \citep{2023A&A...672A...2E, 2023MNRAS.518.1260S} of  a strong lens system 2M1134--2103 are independently estimated and separately published.


We focus on how to combine these estimates available  in the literature to estimate $H_0$ via a meta-analysis \citep[Chapter~5.6]{gelman2013bayesian}. The meta-analysis is built on the estimates of time delays and those of Fermat potential differences, instead of modeling time series data and high-resolution imaging data from scratch. \textcolor{black}{The meta-analysis will be particularly useful in the upcoming era of the Rubin Observatory because the telescope will monitor thousands of strong lens systems  for ten years \citep{2010MNRAS.405.2579O}, and  automated inferential tools will produce numerous time delay and Fermat potential difference estimates \citep{shajib2019is, 2020A&A...640A.105M, 2023MNRAS.522.1323L, 2023MNRAS.518.1260S, 2023A&A...672A...2E}.}  We adopt a Student's $t$ measurement error model by assuming that the input estimates are measured  around  unknown true values  with  Student's $t$ noises.  The scale \textcolor{black}{of the $t$-distribution} is set to the given standard error of the estimate, and thus it is assumed to be fully known.  This heavy-tailed error assumption makes the model  robust to potential outliers. \textcolor{black}{This aspect is  of great importance in practice because there are not many strong lens systems that are currently used to infer $H_0$, such as seven lens systems in \cite{birrer2020tdcosmo}. When the data size is small, any  single biased input can substantially influence the resulting $H_0$ estimate, and Student's $t$ error can effectively prevent the unwanted impact of possibly wrong inputs.} Next, we incorporate physical equations into the model to  \textcolor{black}{estimate} the unknown true values, additionally accounting for more source of uncertainty such as an effect of the mass along the line of sight between the lens and the observer. The model ends up with $(2+K)$  parameters, where $K$ is the number of lens systems in the data. Weakly informative Uniform and Cauchy priors are assumed on these $(2+K)$ model parameters. An R package $\texttt{h0}$ fits the model via  Metropolis-Hastings within Gibbs sampling \citep{metropolis1953equation, geman1984stochastic, tierney1994markov}. It takes about 2,900 seconds on average to draw 10,000 posterior samples when the data are composed of 90 input pairs (time delay and Fermat potential difference) of 30 quadruply-lensed systems.

%



Our numerical studies based on both simulated and realistic data show that the proposed meta-analysis can produce an accurate $H_0$ estimate \textcolor{black}{even in the presence of outlying inputs. The first numerical study is based on a simulated data set publicly available}  from the Time Delay Lens Modeling Challenge \citep{tdlmc2020result}, a blind data analytic competition held from 2018 to 2019.  Specifically, we check how \textcolor{black}{the $H_0$ estimate changes when we manipulate more and more  input data to be wrong or outliers. It shows that the $H_0$ estimate is robust even when 30\% of the 40 inputs from 16 lens systems are modified to be outliers. The second numerical study is based on realistic data of 90 input pairs from 30  quad-lens systems analyzed by the STRIDE science  collaboration \citep{2023MNRAS.518.1260S}. This simulation also confirms that the meta-analysis can recover their underlying cosmological parameters  in a robust manner when 30\% of data are manipulated to be outliers. Finally, we apply the meta-analysis to  three strong lens systems, and estimate $H_0$ by \textcolor{black}{$75.81 \pm 6.82$}. This estimate is consistent to previous $H_0$ estimates under time delay cosmography,  for example,  73.3$^{+1.7}_{-1.8}$  in \cite{wong2020}, 74.2 $^{+2.7}_{-3.0}$  in \cite{shajib2020}, and 74.5 $^{+5.6}_{-6.1}$ in \cite{birrer2020tdcosmo}, not to mention the estimates from the early and late Universe measurements in tension.}  

The rest of this article is organized as follows. We describe details of time delay cosmography in Section~\ref{sec2} \textcolor{black}{and outlines the proposed meta-analysis in Section~\ref{sec:meta}. Modeling assumptions and model fitting procedures  via maximum likelihood estimation and Bayesian posterior sampling appear in Section~\ref{sec3}}.  We investigate the performance of the  meta-analysis in two simulation settings in Sections~\ref{sec5.1tdlmc} and \ref{sec5.2strides}, and \textcolor{black}{explain how we obtain the $H_0$ estimate from the three lens systems} in Section~\ref{sec5.3real}. We discuss limitations and future direction of the meta  analysis in Section~\ref{sec5}.


\section{\textcolor{black}{Overview of} Time Delay Cosmography}\label{sec2}
Time delay cosmography infers $H_0$ using the information about the additional travel distance of the light caused by strong gravitational lensing, \textcolor{black}{as visualized in the left panel of Figure~\ref{fig1lensing}}. The  following physical equation plays a key role in describing this additional travel distance \citep{ref1964h0}:
\begin{equation}\label{equ1}
c\Delta_{ijk}=D_{\Delta}(H_0, z_k, \Omega) \phi_{ijk},
\end{equation}
where $c$ denotes the speed of light that is about \textcolor{black}{$8.39 \times 10^{-7}$ megaparsecs per day (that is, about $2.59\times 10^{10}$ kilometers per day)}, and $\Delta_{ijk}$ is the time delay in days between lensed images $i$ and $j$ of quasar~$k$ ($k=1, 2, \ldots, K$). Intuitively, the multiplication of these two quantities, $c\Delta_{ijk}$, on the left-hand side of Eqn~\eqref{equ1} represents the additional travel distance caused by strong gravitational lensing in  that the light speed (km/day) is multiplied by the additional travel time (day). 

\begin{figure}[b!]
\includegraphics[scale = 0.4]{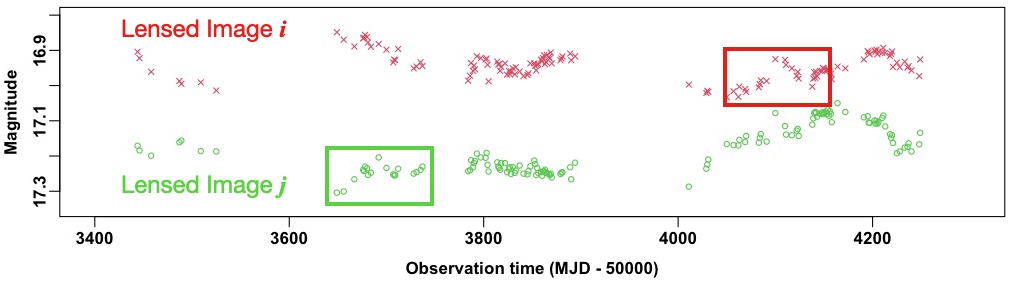}
\caption{\textcolor{black}{Time series data of brightness obtained by measuring brightness of doubly-lensed images of quasar Q0957+561 in $g$-band over time \citep{2012A&A...540A.132S}. Magnitude is an astronomical measure of brightness,  observation time is in days, and MJD represents modified Julian day. A time delay estimate between two lensed images can be obtained by identifying a time lag between similar fluctuation patterns appearing in both time series. }}
\label{timedelays}
\end{figure}

\textcolor{black}{Time delays are estimated from multiple time series data of brightness of lensed images. These data are obtained by measuring brightness of multiply lensed images of the same source in the sky, such as the two bright images around the center in the right panel of Figure~\ref{fig1lensing}. For instance, Figure~\ref{timedelays} exhibits two time series data of brightness (magnitude) for doubly lensed images ($i$ and $j$) of quasar Q0957+561, observed in \textcolor{black}{a $g$-band optical filter based on Sloan Digital Sky Survey photometric system} \citep{2012A&A...540A.132S}. Here, it is not difficult to identify similar fluctuation patterns appearing in both time series, such as those appearing in two rectangles, with about 400-day-long time lag. A model-based approach to time delay estimation treats this time lag as an unknown parameter to be estimated. For example, \cite{2020AJ....160..265H} estimate the time delay of this data set as 413.392  days by adopting a damped random walk model \citep{kelly2009variations} to describe stochastic variability of the time series data.}

The right-hand side of Eqn~\eqref{equ1} re-expresses the  additional travel distance  under the Einstein's \textcolor{black}{general relativity}, accounting for light photons traveling in the curved space and time caused by the strong gravitational field of the intervening galaxy. The notation $z_k=\{z_{sk}, z_{dk}\}$ indicates a vector for two redshifts measuring how fast the $k$-th quasar ($z_{sk}$) and lensing galaxy ($z_{dk}$) are moving away from the observer; the subscripts $s$ and $d$ denote `source (quasar)' and `deflector (lens)', respectively. These redshifts can be accurately measured \textcolor{black}{via spectroscopic data}, and thus we assume that the vector $z_k$ is fully known for all~$k$. The next notation $\Omega=\{\Omega_{\textrm{m}}, \Omega_{\Lambda}\}$ represents a vector for two cosmological parameters, the present-day dark matter density $\Omega_{\textrm{m}}$ and dark energy density $\Omega_{\Lambda}$. Since their sum is one under the standard cosmological model, that is, $\Omega_{\textrm{m}}+\Omega_{\Lambda}=1$, we consider $\Omega_{\textrm{m}}$ as the only unknown parameter in $\Omega$. Hereafter, we use $\Omega$ or $\Omega_{\textrm{m}}$ exchangeably.

\begin{figure}[b!]
\includegraphics[scale = 0.4]{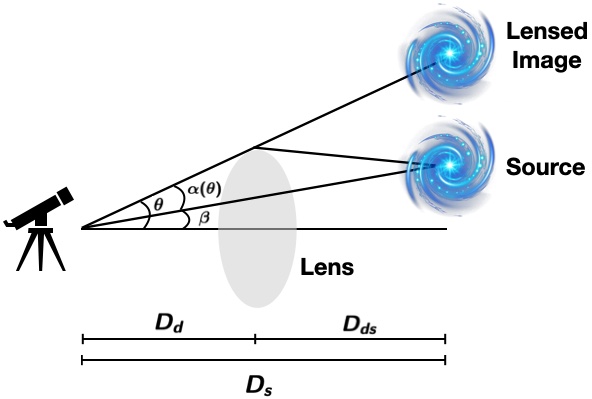}
\caption{\textcolor{black}{Angular distances and positions. Angular diameter distance from the observer to the deflector (lens) is $D_d$, that from the observer to the source is $D_s$, and that from the deflector to the source is $D_{ds}$. Angular position of the lensed image is $\theta$, and that of the source is $\beta$. The scaled deflection angle of the lensed image at $\theta$ is denoted by $\alpha(\theta)$, satisfying the lens equation $\beta=\theta-\alpha(\theta)$.}}
\label{angular}
\end{figure}

The notation  $D_{\Delta}(H_0, z_k, \Omega)$ denotes the time delay distance in the unit of megaparsec (Mpc) defined as a ratio of three angular diameter distances from the observer to the deflector ($D_d$), from the observer to the source ($D_s$), and from the deflector to the source ($D_{ds}$). \textcolor{black}{Figure~\ref{angular} visualizes these angular diameter distances.}  Specifically, these three  distances are deterministic functions of $H_0, z_k$, and $\Omega$:
\begin{align}
\begin{aligned}\label{angulardist}
D_d(H_0, z_k, \Omega)&=\frac{c\int_0^{z_{dk}} \frac{du}{W(u; \Omega)}}{(1+z_{dk})H_0},~~~~~D_s(H_0, z_k, \Omega)=\frac{c\int_0^{z_{sk}} \frac{du}{W(u; \Omega)}}{(1+z_{sk})H_0},~\textrm{and}\\
D_{ds}(H_0, z_k, \Omega)&=\frac{c}{(1+z_{sk})H_0}\left(\int_0^{z_{sk}}\frac{du}{W(u; \Omega)}-\int_0^{z_{dk}}\frac{du}{W(u; \Omega)}\right)
\end{aligned}
\end{align}
The notation  $W(u; \Omega)$ denotes  a deterministic function of $u$ given the cosmological parameters, $\Omega_\textrm{m}$ and $\Omega_\Lambda$. Specifically,
$W(u; \Omega)=((1+u)^3\Omega_{\textrm{m}}+\Omega_{\Lambda})^{0.5}$ under the standard cosmological  model; see \cite{hogg1999distance} for details of cosmological distance measures.  Then the time delay distance of lens system $k$ is defined by these angular diameter distances:
\begin{align}
\begin{aligned}\label{timedelaydist}
D_{\Delta}(H_0, z_k, \Omega)&=(1+z_{dk}) \frac{D_d(H_0, z_k, \Omega)D_s(H_0, z_k, \Omega)}{D_{ds}(H_0, z_k,   \Omega)}\\
&\textcolor{black}{=\frac{c\left(\int_0^{z_{dk}} \frac{du}{W(u; \Omega)}\right)\left(\int_0^{z_{sk}} \frac{du}{W(u; \Omega)}\right)}{H_0\left(\int_0^{z_{sk}}\frac{du}{W(u; \Omega)}-\int_0^{z_{dk}}\frac{du}{W(u; \Omega)}\right)}}.
\end{aligned}
\end{align}
\textcolor{black}{Clearly, the time delay distance $D_{\Delta}(H_0, z_k, \Omega)$ is inversely proportional to $H_0$,  making it useful for the Hubble constant estimation \citep{suyu2017holicow, 2022A&ARv..30....8T}.}




The last \textcolor{black}{quantity} $\phi_{ijk}$ in Eqn~\eqref{equ1} is the difference in Fermat potential that light rays of lensed images $i$ and $j$ of the $k$-th lens system pass through \citep[Section 2.2]{schneider2006}. This represents the different extents to the light deflection and to scaled gravitational potential at two different regions of the gravitational field. Specifically this Fermat potential difference is  defined as 
\begin{equation}\label{lenspotential}
\phi_{ijk} = \phi(\theta_{ik}, \theta_{jk}, \beta_k) =\frac{1}{2}(\theta_{ik} - \beta_{k})^2 -\psi(\theta_{ik}) - \left(\frac{1}{2}(\theta_{jk} - \beta_{k})^2 -\psi(\theta_{jk})\right).
\end{equation}
Here $\theta_{ik}$ and $\theta_{jk}$ are the apparent \textcolor{black}{angular} positions  of lensed images $i$ and $j$ of source $k$ on the sky, respectively, \textcolor{black}{and each is a vector of length two for angular coordinates (ascension and declination)}. The notation $\beta_{k}$ is the apparent \textcolor{black}{angular} position of source $k$ that could have been observed without a lens.   In practice, we cannot observe $\beta_{k}$ because the \textcolor{black}{intervening} lensing galaxy  \textcolor{black}{blocks} the line of sight between the source and the observer.  \textcolor{black}{The difference between $\theta_{ik}$ and $\beta_{k}$ in Eqn~\eqref{lenspotential} is called the scaled deflection angle of the lensed image $i$ of source $k$, which is typically denoted by $\alpha(\theta_{ik})$ in the literature. This relationship forms a so-called lens equation, $\beta_k=\theta_{ik}-\alpha(\theta_{ik})$ for any lensed image $i$. These angular positions and deflection angle are illustrated in Figure~\ref{angular} with  one lensed image.} Lastly, $\psi(\theta_{ik})$ and $\psi(\theta_{jk})$ are scaled gravitational potential of the lensing galaxy (or simply called lens potential) at the positions of lensed images, $\theta_{ik}$ and $\theta_{jk}$, respectively. \textcolor{black}{The lens potential function must satisfy two conditions. First, the gradient of each lens potential becomes the scaled deflection angle, that is, $\alpha(\theta_{ik})=\nabla \psi(\theta_{ik})$, so that the gradient of Fermat potential is zero according to the Fermat's principle. Second, one half of the Laplacian becomes the  dimensionless surface mass density $\kappa(\theta_{ik})~(=\nabla^2 \psi(\theta_{ik})/2$). This dimensionless density can also be expressed as $\kappa(\theta_{ik})=\Sigma(\theta_{ik})/\Sigma_{\textrm{cr}}$, where $\Sigma(\theta_{ik})$ is the surface mass density function and $\Sigma_{\textrm{cr}}$ is the critical mass density.}

\textcolor{black}{Thus, the lens modeling starts by adopting a functional form of this surface mass density, $\Sigma(\theta_{ik})$; its integral becomes $\alpha(\theta_{ik})$ and its double-integral is the lens potential $\psi(\theta_{ik})$.  Stellar kinematic information can be incorporated into the model to handle mass-sheet degeneracy, an effect that different lens mass model can produce the same observations   \citep{1988ApJ...327..693G, 2016JCAP...08..020B, 2023A&A...675A..21Y, 2023A&A...673A...9S}.  In addition to this lens mass model, surface brightness is also modeled for each of lens and source. With a set of unknown parameters in these lens mass, lens light, and surface light models, including unknown angular positions, one can simulate imaging data, that is, a model prediction of the imaging data  given the model parameters. Then, a likelihood function of all of the model parameters  can be obtained by an independent Gaussian assumption  on the observed light intensity  in each pixel whose mean is the pixel-wise  model prediction of  light intensity.  See Section 5 of \cite{2022arXiv221010833B} for more details of  this estimation procedure.}

\textcolor{black}{Besides the explicitly stated quantities in Eqn~\eqref{equ1}, astronomers account for the effect of the mass structure along the line of sight between the lens and the observer because  it is known to be an important source of bias and extra uncertainty in $H_0$ estimation \citep[Section 3.2.4]{suyu2010external, 2013A&A...559A..37S, 2013ApJ...766...70S, 2014ApJ...788L..35S, 2014MNRAS.437..600S, 2017MNRAS.467.4220R,  2020MNRAS.498.3241B, 2020MNRAS.498.1406T, 2021JCAP...08..024F, 2023arXiv230203176W, 2022A&ARv..30....8T}.  This effect is  characterized by a quantity called external convergence, denoted by $\kappa_{\text{ext}, k}$ for each lens system $k$, whose value is negative if the line of sight has less dense structure than the  overall density of the Universe and is positive for more dense structure. The time delay distance after accounting for this line-of-sight effect is defined as
\begin{equation}\label{externalconv}
D_{\Delta}^{\textrm{ext}}(H_0, z_k, \Omega)=\frac{D_{\Delta}(H_0, z_k, \Omega)}{1-\kappa_{\text{ext}, k}}.
\end{equation}
Since the Hubble constant is inversely proportional to the time delay distance, this line-of-sight effect is propagated to the Hubble constant as follows.} 
\begin{equation}\label{externalh0}
H_0^{\textrm{ext}}=(1-\kappa_{\textrm{ext}, k})H_0.
\end{equation}
\textcolor{black}{Because of this relationship, a positive value of $\kappa_{\textrm{ext}, k}$ means that $H_0$ will be over-estimated  if we do not  account for the over-dense line-of-sight effect.  Also, the uncertainty of the resulting $H_0^{\textrm{ext}}$ estimate will also be affected by the multiplication factor $(1-\kappa_{\textrm{ext}, k})$. This quantity $\kappa_{\textrm{ext}, k}$ can be  estimated from external data by comparing the number of galaxies in the lens field of interest with the counts of galaxies in  similar reference fields, without knowing the time delay and Fermat potential difference estimates. See \cite{2023arXiv230203176W} for more details of modeling the external convergence.}

\begin{figure}[b!]
\begin{center}
\includegraphics[scale = 0.32]{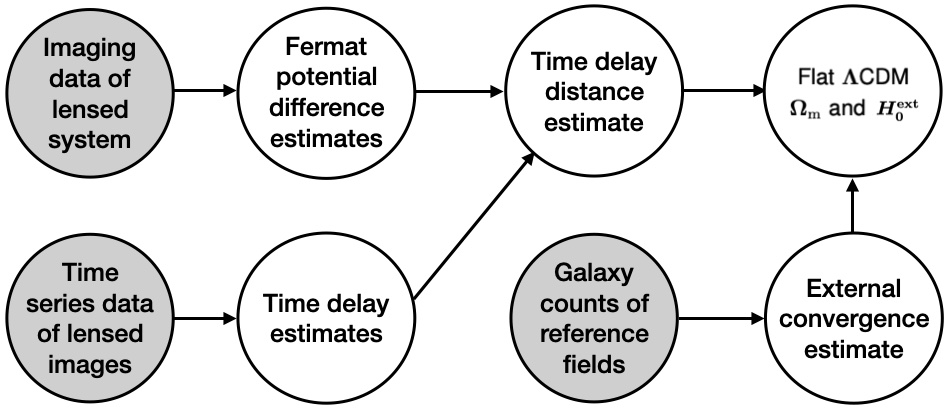}
\caption{This diagram illustrates a series of estimation steps towards the Hubble constant $H_0^{\text{ext}}$. The gray circles indicate the observed data.}
\label{timedelaycosmo}
\end{center}
\end{figure}


\section{Meta-Analysis}\label{sec:meta}
Time delay cosmography is a comprehensive framework to infer $H_0$ \textcolor{black}{based mainly on physical equation in Eqn~\eqref{equ1}}. The equation relates \textcolor{black}{six} quantities, that is, \textcolor{black}{light speed $c$}, time delay $\Delta_{ijk}$, the Hubble constant $H_0$, redshifts $z_k$, cosmological parameter $\Omega$, and Fermat potential difference $\phi_{ijk}$. \textcolor{black}{Among these, two quantities, $c$ and $z_k$ are fixed at known constants.}  The time delays ($\Delta_{ijk}$'s) and Fermat potential differences ($\phi_{ijk}$'s) are independently estimated from multiple time series data of brightness of lensed images and \textcolor{black}{from} high-resolution imaging data, respectively. These estimates can constrain the time delay distance $D_{\Delta}(H_0, z_k, \Omega)$ via Eqn~\eqref{equ1} for all $k$, and then the information about these multiple time delay distances, $D_{\Delta}(H_0, z_1, \Omega)$, $D_{\Delta}(H_0, z_2, \Omega)$, $\ldots$, $D_{\Delta}(H_0, z_K, \Omega)$ can constrain $H_0$ and $\Omega$ in turn. \textcolor{black}{To account for potential bias and extra uncertainty caused by the line-of-sight effect, astronomers incorporate external convergence into the model  and finally estimate $H_0^{\text{ext}}$.} Figure~\ref{timedelaycosmo} illustrates this series of inferential steps  toward $H_0^{\text{ext}}$ under time delay cosmography \textcolor{black}{in a simplified manner}. The circles in gray indicate the observed data.



\textcolor{black}{Estimating $H_0$ via time delay cosmography requires enormous  efforts of more than a hundred experts from various fields of astronomy. 
Also, there are at least five scientific collaboration groups contributing to time delay cosmography;  COSMOGRAIL collaboration \citep{2005A&A...436...25E}; H0LiCOW  collaboration \citep{suyu2017holicow};  STRIDES collaboration \citep{2018MNRAS.481.1041T}; SHARP collaboration \citep{2019MNRAS.490.1743C}; and   TDCOSMO collaboration \citep{millon2020}. These are clear evidence that the Hubble constant estimation via time delay cosmography is not only of great interest in astronomy, but also challenging enough to require enormous human efforts and time.}

\begin{figure}[b!]
\begin{center}
\includegraphics[scale = 0.32]{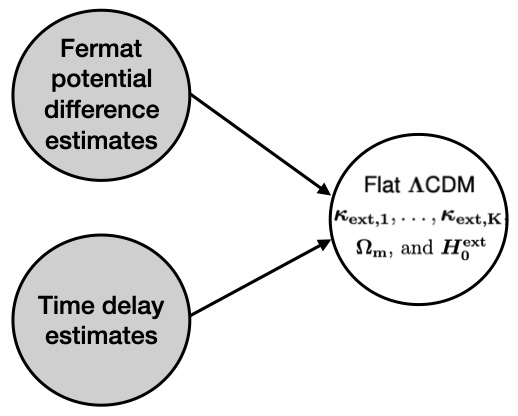}
\caption{This diagram illustrates how the  meta-analysis infers the Hubble constant $H_0$. The gray circles indicate the input data that are pairs of the Fermat potential difference  and time delay estimates. The meta-analysis treats time delay distance $D_{\Delta}(H_0, z_k, \Omega)$ for each lens system $k$ as a deterministic function of $H_0$, $z_k$, and $\Omega$, as its notation indicates, instead of treating each time delay distance as a quantity to be estimated before constraining $H_0$ and $\Omega$. Also, the meta-analysis accounts for the line-of-sight effect by treating external convergences, $\kappa_{\text{ext}, 1}, \ldots, \kappa_{\text{ext}, K}$, as unknown model parameters.}
\label{metanalysis}
\end{center}
\end{figure}

Our motivation is that a meta-analysis can save such humongous efforts by extracting the information about $H_0$ from various estimates separately published in the literature, as the information is commonly embedded in these independent studies.  \textcolor{black}{Also,} since there are not many  strong lens systems \textcolor{black}{currently available} for accurate $H_0$ estimation, for example, with six to eight lens systems \citep{shajib2020, wong2020, birrer2020tdcosmo, 2021MNRAS.501..784D, 2022MNRAS.514.1433W}, it is crucial to prevent outlying inputs from unduly influencing the $H_0$ estimation. 


Thus, we propose a robust meta-analysis based on Student's $t$ error that takes the time delay and Fermat potential difference estimates as inputs, without modeling the time series and imaging data from scratch.  We incorporate the external convergence of each lens system into the model as unknown model parameters, instead of estimating them from external data. Unlike the common practice in time delay cosmography, \textcolor{black}{the meta-analysis does not require any blind analysis between estimations of time delay and Fermat potential difference  because it takes the results of the blind analysis as inputs. Also,} the meta-analysis does not separately estimate time delay distances, $D_{\Delta}(H_0, z_k, \Omega)$'s, before constraining $H_0$ and $\Omega$. Instead, it treats these deterministic functions as a medium to access the information about $H_0$ and $\Omega$ from the inputs. This has an effect of increasing the data size. For example, a quad-lens system $k$ produces at least three pairs of Fermat potential difference and time delay estimates, such as $(\hat{\phi}_{12k}, \hat{\Delta}_{12k})$, $(\hat{\phi}_{13k}, \hat{\Delta}_{13k})$, and $(\hat{\phi}_{14k}, \hat{\Delta}_{14k})$. The meta-analysis treats these as three independent inputs that contain the information about $H_0$, instead of reducing these three pairs to  one time delay distance estimate, $\hat{D}_{\Delta}(H_0, z_k, \Omega)$, to infer $H_0$.  Treating the three paired estimates as three independent observations  does not mean that each pair equally contributes to the $H_0$ estimate because each pair of estimates has different uncertainty levels (standard errors). 
Figure~\ref{metanalysis} illustrates a workflow of the meta-analysis.

\textcolor{black}{One notable limitation of the proposed meta-analysis is that  it considers paired estimates in a quad-lens system as independent observations, even though they are actually dependent within the same lens system. For example, $\hat{\Delta}_{12k}$, $\hat{\Delta}_{13k}$, and $\hat{\Delta}_{14k}$ (or $\hat{\phi}_{12k}$, $\hat{\phi}_{13k}$, and $\hat{\phi}_{14k}$) can have some physical or statistical dependence as they are estimated in the same lens system. Thus, it is more principled to model such dependence via a multivariate Student's $t$ distribution on $\hat{\Delta}_{12k}$, $\hat{\Delta}_{13k}$, and $\hat{\Delta}_{14k}$ and another multivariate Student's $t$ distribution on $\hat{\phi}_{12k}$, $\hat{\phi}_{13k}$, and $\hat{\phi}_{14k}$, if the information about their correlations is available. However, such information is neither easily accessible nor  always available in practice, which is our main motivation to assume the independence across input pairs. Also, the independence assumption has a computational advantage because it comes with a linearly increasing  computation cost in the likelihood computation (linear in the number of input pairs). This computational aspect will become crucial when the number of input pairs considerably increases in the era of the Rubin Observatory.}

\textcolor{black}{Also we note that the proposed method does not take mass-sheet degeneracy \citep{2021A&A...652A...7C, 2023A&A...675A..21Y} into account. This degeneracy means that for any given lens model, there exists a whole family of mass distributions that are perfectly degenerate and fit the lensing data equally well.  Since incorrect mass normalization directly translates into a bias in the inferred value of $H_0$, and the overall mass normalization cannot be uniquely determined from the lensing data alone, additional information is necessary to break this degeneracy, such as the line-of-sight velocity dispersion. However, the current work does not model  line-of-sight velocity dispersion.}


\section{Statistical Modeling and Inference}\label{sec3}


We derive the likelihood function of $H_0$, $\Omega$, \textcolor{black}{and $\kappa_{\text{ext}, k}$'s by combining Eqns~\eqref{equ1} and \eqref{externalconv}, assuming that the time delay and Fermat potential difference constrain $D^{\text{ext}}_{\Delta}(H_0, z_k, \Omega)$, instead of $D_{\Delta}(H_0, z_k, \Omega)$ in Eqn~\eqref{equ1}. That is,} \begin{equation}\label{equ7}
c\Delta_{ijk}=D_{\Delta}^{\textcolor{black}{\text{ext}}}(H_0, z_k, \Omega)\phi_{ijk}=\frac{D_{\Delta}(H_0, z_k, \Omega)\phi_{ijk}}{\textcolor{black}{1-\kappa_{\text{ext}, k}}}.
\end{equation}
Then, the Fermat potential difference can be expressed as
\begin{equation}\label{equ8}
\phi_{ijk}=\textcolor{black}{\frac{(1-\kappa_{\text{ext}, k})c\Delta_{ijk}}{D_{\Delta}(H_0, z_k, \Omega)}}.
\end{equation}
\textcolor{black}{Applying a Gaussian assumption on the Fermat potential difference estimate, as  proposed in an unpublished work of  Marshal et al., we obtain the following Gaussian distribution of the Fermat potential difference estimate centered at the unknown true Fermat potential difference $\phi_{ijk}$ in Eqn~\eqref{equ8}:}
\begin{equation}\label{model1}
\hat{\phi}_{ijk}\mid H_0, \Omega,  \Delta_{ijk}, \textcolor{black}{\alpha_{ijk}, \kappa_{\text{ext}, k}}\sim N\!\left(\frac{\textcolor{black}{(1-\kappa_{\text{ext}, k})}c\Delta_{ijk}}{D_{\Delta}(H_0, z_k, \Omega)},~~ \textcolor{black}{\alpha_{ijk}}\hat{\sigma}^2_{\phi_{ijk}}\right),
\end{equation}
where $\hat{\sigma}_{\phi_{ijk}}$ is  the given standard error of  $\hat{\phi}_{ijk}$, \textcolor{black}{and $\alpha_{ijk}$ is an unknown variance adjustment factor to account for  extra uncertainty beyond the given standard error, which is  called an error-on-the-error approach in particle physics \citep{2019EPJC...79..133C}.  This adjustment factor later transforms  Gaussian measurement error into Student's $t$ measurement error \citep{tak2019robust}. Conditioning on $(H_0, \Omega,  \Delta_{ijk}, \kappa_{\text{ext}, k})$ in Eqn~\eqref{model1} is equivalent to conditioning on $\phi_{ijk}$ because of their deterministic relationship in Eqn~\eqref{equ8}.}  Marshal et al.~fix $\Omega_{\textrm{m}}~\textcolor{black}{(=\Omega~\text{in this work})}$ at 0.3, but  we treat it as an unknown parameter to be estimated.  

Similarly, we assume that the time delay estimate  $\hat{\Delta}_{ijk}$  follows a Gaussian distribution \textcolor{black}{\citep[Section 4.6]{2016JCAP...08..020B}}, whose mean is  the unknown true time delay $\Delta_{ijk}$  and variance is equal to squared standard error , $\hat{\sigma}^2_{\Delta_{ijk}}$, \textcolor{black}{multiplied by the same adjustment factor, $\alpha_{ijk}$}:
\begin{equation}\label{model2}
\hat{\Delta}_{ijk}\mid \Delta_{ijk}, \textcolor{black}{\alpha_{ijk}}\sim N\!\left({\Delta}_{ijk},~~ \textcolor{black}{\alpha_{ijk}}\hat{\sigma}^2_{\Delta_{ijk}}\right).
\end{equation} 

Denoting all  time delays by $\Delta=\{\Delta_{ijk}: i<j~\textrm{and}~ k=1, 2, \ldots, K\}$, \textcolor{black}{all  adjustment factors by $\alpha=\{\alpha_{ijk}: i<j~\textrm{and}~ k=1, 2, \ldots, K\}$, and all external convergences by $\kappa_{\text{ext}}=\{\kappa_{\text{ext}, k}: k=1, 2, \ldots, K\}$}, we express the likelihood function of $H_0, \Omega$, $\Delta$, \textcolor{black}{$\alpha$, and $\kappa_{\text{ext}}$} as a multiplication of joint density functions  of input data pairs, ($\hat{\phi}_{ijk}, \hat{\Delta}_{ijk})$'s, denoted by $f$: 
\begin{align}
\begin{aligned}\label{lik1}
L(H_0, \Omega, \Delta, \textcolor{black}{\alpha, \kappa_{\text{ext}}}) &=\prod_{k=1}^K\prod_{i<j} f(\hat{\phi}_{ijk}, \hat{\Delta}_{ijk}\mid H_0, \Omega, \Delta_{ijk}, \textcolor{black}{\alpha_{ijk}, \kappa_{\text{ext}, k}})\\
&=\prod_{k=1}^K\prod_{i<j} g(\hat{\phi}_{ijk}\mid H_0, \Omega, \Delta_{ijk}, \textcolor{black}{\alpha_{ijk}, \kappa_{\text{ext}, k}})h(\hat{\Delta}_{ijk}\mid \Delta_{ijk}, \textcolor{black}{\alpha_{ijk}}).
\end{aligned}
\end{align}
The \textcolor{black}{functions $g$ and $h$ represent densities of the two Gaussian distributions defined in   Eqns~\eqref{model1} and \eqref{model2}, respectively. The factorization in Eqn~\eqref{lik1} is based on conditional independence assumption between $\hat{\phi}_{ijk}$ and $\hat{\Delta}_{ijk}$  given $\Delta_{ijk}$, and that between $\hat{\Delta}_{ijk}$ and $(H_0, \Omega, \kappa_{\text{ext}, k})$ given $\Delta_{ijk}$. The first conditional independence makes sense because conditioning on $\hat{\Delta}_{ijk}$ in $g$ becomes redundant while the true value $\Delta_{ijk}$ is already in the condition.  The second conditional independence can also be justified because knowing $(H_0, \Omega, \kappa_{\text{ext}, k})$ does not affect the distribution of the time delay estimate once we condition on the true time delay $\Delta_{ijk}$ and $\alpha_{ijk}$.}


Next, we integrate out the true time delays $\Delta$ \textcolor{black}{and  adjustment factors $\alpha$} from the likelihood function in Eqn~\eqref{lik1} \textcolor{black}{after assuming unbounded uniform  distribution on $\Delta_{ijk}$, $\pi_1(\Delta_{ijk})\propto 1$,  and inverse-Gamma(2, 2)  distribution on $\alpha_{ijk}$, $\pi_2(\alpha_{ijk})\propto \alpha_{ijk}^{-3}\exp(-2/\alpha_{ijk})$}. The \textcolor{black}{uniform marginal} distribution is adopted to reflect our lack of knowledge about $\Delta_{ijk}$, and the inverse-Gamma(2, 2) distribution is to transform the Gaussian error to Student's $t$ error \textcolor{black}{with four degrees of freedom}. \textcolor{black}{We fix the value of the degrees of freedom  at a constant, instead of treating it as an unknown parameter to be estimated, because we  adopt the Student's $t$ error as a robust alternative to Gaussian error \citep[Chp.~17.2]{gelman2013bayesian}. Also, we consider four degrees of freedom because this choice produces the heaviest tails among the integer-valued degrees of freedom that make the mean and variance of the $t$-distribution exist.} The marginalization of $\Delta$ \textcolor{black}{and $\alpha$} results in an integrated likelihood function of $H_0$, $\Omega$, \textcolor{black}{and $\kappa_{\text{ext}}$} \citep{berger1999integrated}:
\begin{align}
\begin{aligned}\label{lik2}
L(H_0, \Omega, \textcolor{black}{\kappa_{\text{ext}}})&=\int L(H_0, \Omega, \Delta, \textcolor{black}{\alpha, \kappa_{\text{ext}}})\prod_{k=1}^K\prod_{i<j} \pi_1(\Delta_{ijk})\textcolor{black}{\pi_2(\alpha_{ijk})}~d\Delta_{ijk}~\textcolor{black}{d\alpha_{ijk}}\\
&=\prod_{k=1}^K\prod_{i<j}  \int g(\hat{\phi}_{ijk}\mid H_0, \Omega, \Delta_{ijk}, \textcolor{black}{\alpha_{ijk}, \kappa_{\text{ext}, k}})h(\hat{\Delta}_{ijk}\mid \Delta_{ijk}, \textcolor{black}{\alpha_{ijk}})\\
&~~~~~~~~~~~~~~~~~~~~\times\pi_1(\Delta_{ijk})\textcolor{black}{\pi_2(\alpha_{ijk})}~d\Delta_{ijk}~\textcolor{black}{d\alpha_{ijk}}\\
&=\prod_{k=1}^K\prod_{i<j} p(\hat{\phi}_{ijk}\mid H_0, \Omega, \textcolor{black}{\kappa_{\text{ext}, k}}, \hat{\Delta}_{ijk}, ).
\end{aligned}
\end{align}
The density function $p$  in Eqn~\eqref{lik2} is a \textcolor{black}{Student's $t$} density function  of  $\hat{\phi}_{ijk}$. \textcolor{black}{To be specific,}
\begin{equation}\label{model3}
\hat{\phi}_{ijk}\mid H_0, \Omega, \textcolor{black}{\kappa_{\text{ext}, k}}, \hat{\Delta}_{ijk} \sim t\!\left(\frac{\textcolor{black}{(1-\kappa_{\text{ext}, k})}c\hat{\Delta}_{ijk}}{D_{\Delta}(H_0, z_k, \Omega)},~~ \sqrt{\frac{\textcolor{black}{(1-\kappa_{\text{ext}, k})^2}c^2\hat{\sigma}^2_{\Delta_{ijk}}}{D_{\Delta}(H_0, z_k, \Omega)^2}+\hat{\sigma}^2_{\phi_{ijk}}}~~\right).
\end{equation}
The notation $t(a, b)$ represent the Student's $t$ distribution with location parameter $a$ and scale parameter $b$.

For a Bayesian inference, we additionally set up a joint prior distribution of the unknown parameters, $H_0$, $\Omega$,  and $\kappa_{\text{ext}}$, and derive their \textcolor{black}{joint} posterior density function up to a constant multiplication.  \textcolor{black}{As for $H_0$ and  $\Omega$}, the most common choice in the literature is to put a jointly uniform prior on \textcolor{black}{them} to reflect the lack of knowledge about their true values. For example,  Uniform(0, 150) prior on $H_0$, and independently a Uniform$(0.05,~ 0.5)$ prior on $\Omega_{\textrm{m}}$  are the two common choices in the literature  \citep{bonvin2016he, suyu2017holicow, 2017MNRAS.465.4895W, 2019MNRAS.490.1743C, birrer2019sdss, wong2020,  birrer2020tdcosmo, shajib2020, 2020MNRAS.498.1440R, millon2020}. \textcolor{black}{Such non-informative uniform prior distributions are suitable for the current practice of time delay cosmography. This is because not many lens systems are being used to infer $H_0$, and thus informative prior distributions can dominate the resulting posterior inference.} \textcolor{black}{When it comes to $\kappa_{\text{ext}}$, we adopt an independent Cauchy prior distribution with scale 0.025 for each $\kappa_{\text{ext}, k}$, considering that $N(0,~ 0.025)$ is used as a  reasonable simulation assumption for $\kappa_{\text{ext}, k}$ in \cite{tdlmc2020result} and that the Cauchy distribution can cover wider regions than $N(0,~ 0.025)$. Assuming independent external convergences across lens systems is not uncommon in practice \citep{birrer2020tdcosmo}.} We express the resulting joint posterior density of $H_0,  \Omega$, and \textcolor{black}{and $\kappa_{\text{ext}}$} as
\begin{equation}\label{posterior1}
\pi\!\left(H_0, \Omega, \textcolor{black}{\kappa_{\text{ext}}} \mid D\right)\propto L(H_0, \Omega, \textcolor{black}{\kappa_{\text{ext}}})h(H_0, \Omega, \textcolor{black}{\kappa_{\text{ext}}}),
\end{equation}
where $D$ denotes a set of all pairs of the time delay and Fermat potential difference estimates, and $h$ is a joint  prior density function. The posterior distribution is proper because the joint  prior distribution is proper \citep{hobert1996propriety, tak2018how}.

To sample the joint posterior distribution in Eqn~\eqref{posterior1}, \textcolor{black}{the R package  \texttt{h0} adopts  a Metropolis-Hastings within Gibbs sampler \citep[Section 2.4]{tierney1994markov}. It uses a Gaussian proposal distribution for $H_0$ whose proposal scale is adjusted to achieve about 40\% of acceptance rate. The proposals of the other model parameters are independently drawn from their prior distributions \citep[Section 2.3.3]{tierney1994markov}; that is, the proposal distribution for $\Omega$ is Uniform(0.05,~ 0.5), that for each $\kappa_{\text{ext}, k}$ is Cauchy with scale 0.025. We implement five Markov chains each for 10,000 iterations whose initial values are spread throughout the parameter space. Specifically, we set five initial values of $H_0$ to five evenly-spaced values between 0.01 and 150, while the initial values of $\Omega$ and $\kappa_{\text{ext}, k}$'s are randomly drawn from their prior distributions.  For each Markov chain, the first half of the iterations is discarded as burn-in. We combine these five Markov chains  to make a posterior inference after checking the convergence via Gelman-Rubin diagnostic statistic \citep{gelman1992inference}. The R package \texttt{h0} also has a function to conduct a posterior predictive check to see whether there is  evidence that the model does not describe the data well  \citep[][Chapter~6]{10.1214/aos/1176346785, gelman2013bayesian}. In case several modes are identified from the five runs, the R package \texttt{h0} provides an option to replace the Metorpolis update for $H_0$ with  repelling-attracting Metropolis update \citep{tak2018ram}, which enables the Markov chain to jump between local modes frequently. The proposal scale of repelling-attracting Metropolis is adjusted to achieve at least 10\% of acceptance rate.}

\section{\textcolor{black}{Simulated Data Analyses}}

The main goal of \textcolor{black}{two simulation studies in this section}  is to investigate how  \textcolor{black}{robustly} the proposed meta-analysis can infer the true value of $H_0$  \textcolor{black}{in the presence of outlying inputs. For this purpose, we manipulate several inputs to be outliers. We compare the effects of these manipulated inputs on the proposed robust meta-analysis to those on a non-robust meta-analysis equipped with Gaussian error using three criteria; bias in percentage defined as $100(\hat{H}_0-H^{\text{true}}_0)/H^{\text{true}}_0$, coefficient of variation in percentage calculated by $100\hat{\sigma}_{{H}_0}/H^{\text{true}}_0$, and root mean square error computed by the square root of $(\hat{H}_0-H^{\text{true}}_0)^2+\hat{\sigma}_{{H}_0}^2$.  Here, we set $\hat{H}_0$ to the posterior mean and $\hat{\sigma}_{{H}_0}$ to the posterior standard deviation. In the time delay cosmography literature, bias and coefficient of variation are called accuracy and precision, respectively, and are typically reported in percentage. We report CPU times  measured} on a laptop equipped with 2.4GHz Intel quad-core i5 and \textcolor{black}{16} Gb RAM.



\subsection{\textcolor{black}{Simulation Study I}}\label{sec5.1tdlmc}








We borrow the simulation setting of the Time Delay Lens Modeling Challenge \citep{tdlmc2020result}.  It was a  data analytic competition to improve existing tools to analyze high-resolution imaging data of lens systems and to encourage the development of new methods for the  $H_0$ estimation.  The competition had three stages with increasing difficulty in analyzing imaging data. At each stage, the organizers provided  imaging data of 16 lens systems, simulated under the standard cosmological model  and various parameter values including a specific value of $H_0$. They also accounted for  realistic components as well in simulating the  imaging data, for example, by using realistic galaxy images obtained from the  Hubble Space Telescope to simulate  surface brightness.  After the submission deadline of the competition, the blinded values of $H_0$ and the other  true parameter values, such as  time delays ($\Delta_{ijk}$'s) and time delay distances ($D_{\Delta}(H_0, z_k, \Omega)$'s) were completely disclosed to the public. Based on these, it is straightforward to derive the true values of Fermat potential differences ($\phi_{ijk}$'s) via Eqn~\eqref{equ1}.

Our simulation study \textcolor{black}{borrows the setting of} the second stage of the TDLMC. The true values of time delays, Fermat potential differences, and redshifts of 16 lens systems (12 quad-lens systems and 4 double-lens systems) are tabulated in \textcolor{black}{Table~1 of the supplementary materials}. The generative true value of $H_0$ is 66.643, which is the same for all these 16 lens systems.



To investigate \textcolor{black}{the impact of outlying inputs on the $H_0$ estimation, we first prepare an default data set by fixing  the true values at estimates (no bias), that is, $\hat{\Delta}_{ijk}=\Delta_{ijk}$ and $\hat{\phi}_{ijk}=\phi_{ijk}$.} The standard errors of these estimates  are \textcolor{black}{fixed at  3\% of the true values to achieve  3\%  coefficient of variation, that is, $\hat{\sigma}_{\Delta_{ijk}}=0.03 \vert\Delta_{ijk}\vert$ and  $\hat{\sigma}_{\phi_{ijk}}=0.03 \vert \phi_{ijk}\vert$. This is because the community expects state-of-the-art time delay analyses to produce  bias less than 1\% and coefficient of variation less than 3\% for the upcoming large-scale survey by the Rubin Observatory Legacy Survey of  Space of Time \citep{liao2015tdc}. Next, we manipulate several time delay inputs of this default data set to make three cases. The first case contains four manipulated time delay inputs out of 40 (10\%); the second case modifies additional four  inputs (eight in total, 20\%); and the final case changes  another four inputs (12 in total, 30\%). \textcolor{black}{Our intention is to show how the resulting posterior distribution of $H_0$ changes as the proportion of anomalous inputs increases in a linear fashion, for example, 0\%, 10\%, 20\%, and 30\%. The chosen proportions do not imply that realistic data in time delay cosmography actually suffer from such large proportions of outliers. To produce outliers in each case,} we add constants that are 10 times larger than standard errors to the chosen inputs.  \textcolor{black}{For example, the  inputs to be manipulated are $\hat{\Delta}_{21k}$ for $k=1, 2, \ldots, 12$ in  Table~1 of the supplementary materials, and}  the first manipulated input is $\hat{\Delta}_{211}=\Delta_{211}+10\hat{\sigma}_{\Delta_{211}}=\Delta_{211}+0.3\vert \Delta_{211} \vert$.}

\begin{figure}[t!]
\begin{center}
\includegraphics[scale = 0.42]{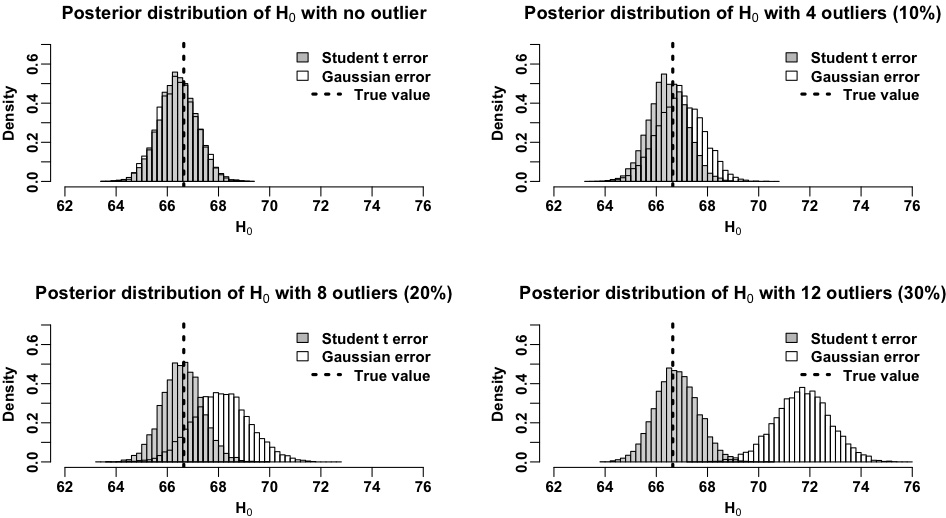}
\caption{The resulting marginal posterior distributions of $H_0$ when there is no outlier (top-left), 4 outliers (top-right),  8 outliers (bottom-left), and 12 outliers (bottom-right). The vertical dashed line in each panel represents the generative true value of $H_0$, 66.643. It shows that the meta-analysis with Student's $t$ error is robust to outlying inputs  unlike the one with Gaussian error.}
\label{ex1res1}
\end{center}
\end{figure}

The resulting posterior distributions of $H_0$ under four cases are displayed  in Figure~\ref{ex1res1}.  In each panel, the gray histogram represents the posterior distribution of $H_0$ obtained with Student's $t$ error, and the white histogram is the distribution obtained with Gaussian error. The true value of $H_0$ is denoted by the vertical dashed line. Without any outlier, both types of error produce almost identical fits, as shown in the top-left panel. However, as we manipulate more and more input time delays, the posterior distributions of $H_0$ under Gaussian error start deviating from the true value with wider spread, while those under Student's $t$ error do not change much. \textcolor{black}{This is because the Student's $t$ error automatically puts less weight on the manipulated inputs in the likelihood computation, while these inputs dominate the likelihood calculation under the Gaussian error.}

\begin{table}[b!]
\caption{\textcolor{black}{Numerical summaries}  obtained by fitting the meta-analysis with either Student's $t$ error or Gaussian error. \textcolor{black}{We report} the posterior means (Post.~mean), posterior standard deviations (sd),  bias in percentage, coefficient of variation (CV) in percentage,  and root mean square error (RMSE). The true value of $H_0$ is 66.643.}
\begin{center}
{\footnotesize\begin{tabular}{cccccc}
\hline
& &No outliers & 4 outliers (10\%)   & 8 outliers (20\%) & 12 outliers (30\%)\\ 
\hline
\multirow{2}{*}{Post.~mean (sd)} & Gaussian& 66.393 (0.761) &  66.933 (0.887)& 68.108 (1.115)& 71.715 (1.086)\\ 
&Student's $t$ & 66.406 (0.730)  &  66.461 (0.739)& 66.581  (0.783)& 66.795  (0.834)\\ 
&&&&&\\
\multirow{2}{*}{Bias (\%)}&Gaussian & 0.375 &   0.435& 2.199 & 7.611\\ 
& Student's $t$  & 0.355 &  0.272 & 0.093 & 0.228\\ 

&&&&&\\
\multirow{2}{*}{CV (\%)} &Gaussian & 1.109 &  1.330 & 1.673& 1.630\\ 
& Student's $t$ &  1.095  &  1.109 & 1.175& 1.251\\ 
&&&&&\\
\multirow{2}{*}{RMSE} &Gaussian & 0.801 &  0.933 & 1.841& 5.187\\ 
& Student's $t$ &  0.768  &  0.761 & 0.786 & 0.848\\ 
\hline
\end{tabular}}\label{table2}
\end{center}
\end{table}

\textcolor{black}{Table~\ref{table2} numerically summarizes the model fits.   We use the posterior mean as an $H_0$ estimate and posterior standard deviation as an $1\sigma$ uncertainty to be comparable with standard errors in existing works. The $H_0$ estimates under Student's $t$ error remain almost the same near the true value of $H_0$  in all cases, while those under Gaussian error change noticeably. These changes are quantified via bias, coefficient of variation, and root mean square error.  For example, when 12 time delay inputs (30\%) are manipulated, the bias  under Gaussian error is 7.611\%, but that under Student's $t$ error is only 0.228\%. The $1\sigma$ uncertainty is also larger under Gaussian error; the coefficient of variation under Gaussian error is 1.630\%  and that under Student's $t$ error is 1.251\%. Its difference is not as noticeable as bias. The root mean square error under Gaussian error is more than five times larger than that under Student's $t$ error (5.187 versus 0.848) mostly due to the contribution of bias.}

As for the computational cost of each simulation, it takes  431 seconds on average to \textcolor{black}{implement 10,000 iterations for each Markov chain. The average Gelman-Rubin diagnostic statistic for $H_0$ is 1.0045 and the largest value for this diagnostic statistic is 1.0135 in  all cases. The average effective sample size for $H_0$ is about 1,200 out of 25,000 posterior samples in each case.  The posterior predictive checks do not show evidence that the model fails to describe the data well; see \textcolor{black}{Figure~1 of the second supplementary material} for more details.}

\subsection{\textcolor{black}{Simulated Study II}}\label{sec5.2strides}

\begin{figure}[b!]
\begin{center}
\includegraphics[scale = 0.21]{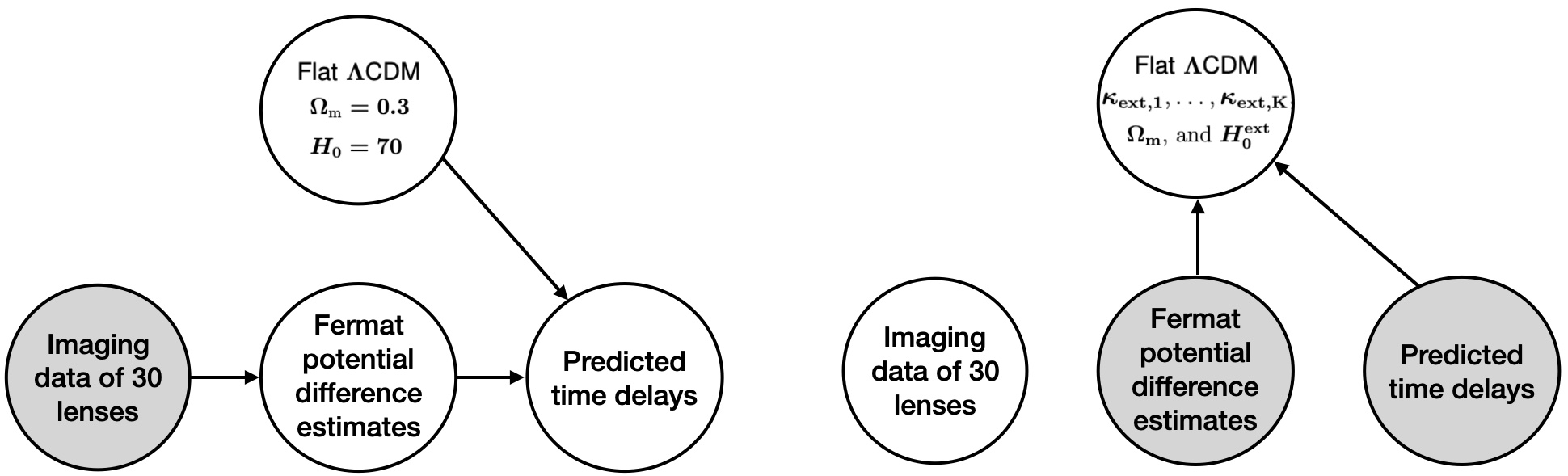}
\caption{The left panel shows the workflow of the most recent work of the STRIDES collaboration \citep{2023MNRAS.518.1260S}. They analyze 31 quad-lens systems using an automated uniform lens modeling approach, reporting estimates of the Fermat potential differences from 30 quad-lens systems (out of 31). Given these estiamtes and fixed cosmological parameters under the standard cosmological model ($\Lambda$CDM), they predict time delays. The right panel shows the workflow of the proposed meta-analysis. We treat their Fermat potential difference estimates and time delay estimates as  inputs to the proposed meta-analysis to see whether the meta-analysis can accurately trace back to the fixed cosmological parameters including $H_0$.}
\label{workflow}
\end{center}
\end{figure}

The most recent work of the STRIDES collaboration \citep{2023MNRAS.518.1260S} analyzes 31 quadruply lensed systems via an automated uniform lens modeling approach \textcolor{black}{that has been improved from}  \cite{shajib2019is}. \textcolor{black}{They estimate} Fermat potential differences of 30 quad-lens systems (out of 31), and predict the corresponding time delays under the standard cosmological model with $\Omega=0.3$ and $H_0=70$. \textcolor{black}{The external convergence  is not modeled in the work of \cite{2023MNRAS.518.1260S}, that is, $\kappa_{\textrm{ext}, k}=0$ for all $k$.} The left panel of Figure~\ref{workflow} shows their workflow. The observed data are colored in gray. The   Fermat potential difference estimates and predicted time delays are listed in Table~8 of \cite{2023MNRAS.518.1260S}. 


We take advantage of their real-data analyses to see whether the proposed meta-analysis can accurately trace back to the true value of $H_0$ given their Fermat potential difference estimates and \textcolor{black}{predicted} time delays. \textcolor{black}{To make the simulation setting more realistic, we treat the  external convergence in each lens system as an unknown model parameter to be estimated}.  The right panel of Figure~\ref{workflow} outlines this work. The gray circles indicate the inputs of the meta-analysis. This setting is ideal for evaluating the performance of the meta-analysis because the STRIDES collaboration has encoded the information about the  cosmological parameters \textcolor{black}{under the standard cosmology, such as $\Omega_{\textrm{m}}$ and $H_0$, into the predicted} time delays. 


\begin{figure}[b!]
\begin{center}
\includegraphics[scale = 0.42]{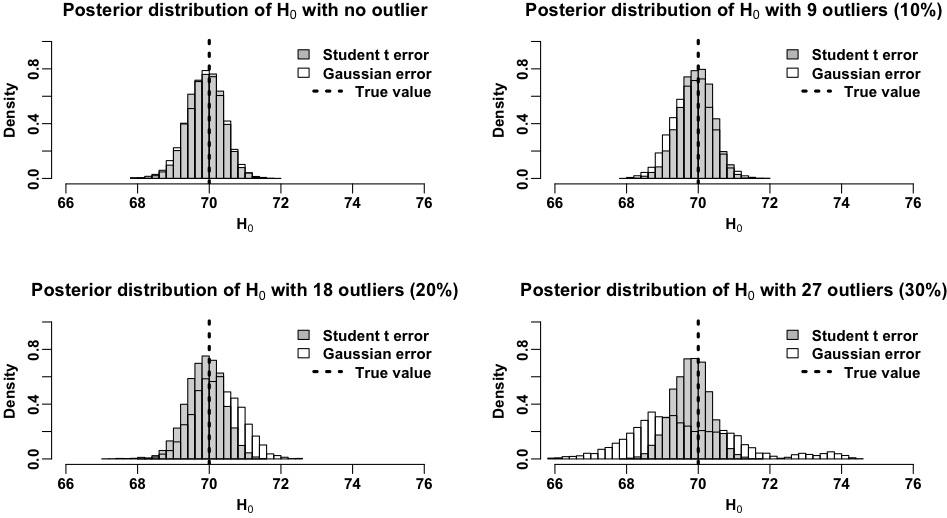}
\caption{The resulting marginal posterior distributions of $H_0$ when there is no outlier (top-left), 9 outliers (top-right),  18 outliers (bottom-left), and 27 outliers (bottom-right). The vertical dashed line in each panel represents the generative true value of $H_0$, 70. It shows that the meta-analysis with Student's $t$ error is robust to outlying inputs. The posterior distributions of Gaussian error starts deviating from the true value of $H_0$ when there are 18 outliers (20\%), ending up with multiple modes in the presence of 27 outliers (30\%).}
\label{ex2res}
\end{center}
\end{figure}

\textcolor{black}{The default data set takes} these  Fermat potential difference estimates and predicted time delays  as inputs. As for  uncertainties of these estimates, the meta-analysis only takes a single-number uncertainty (standard error). However, the reported uncertainties in Table~8 of \cite{2023MNRAS.518.1260S} are 16\% and 84\% percentiles of a posterior distribution, and these percentiles are not always symmetric around each estimate. For example, one of the predicted time delays and its 68\% uncertainty are reported as  $-100^{+4.7}_{-3.0}$. To be conservative, we take the larger distance from the estimate to one of the percentiles as the input uncertainty for the meta-analysis. \textcolor{black}{That means,} we set the standard error of the predicted time delay  to $4.7$ when  $-100^{+4.7}_{-3.0}$ is reported in  Table~8 of \cite{2023MNRAS.518.1260S}. \textcolor{black}{The resulting standard error still satisfies the 3\% coefficient of variation level that the community expects; the median coefficient of variation in percentage is $3.018\%$ for the time delay inputs $(100\hat{\sigma}_{\Delta}/\hat{\Delta})$, and 3.030\% for the Fermat potential difference inputs $(100\hat{\sigma}_{\phi}/\hat{\phi})$.}

\textcolor{black}{To investigate the sensitivity of the meta-analysis to outlying inputs, we set three other data sets by manipulating 9 (10\%), 18 (20\%), and 27 (30\%) time delay inputs out of 90, respectively, as done in the previous simulation study.  The inputs to be manipulated are $\Delta t_{ABk}$ for $k=1, 2, \ldots, 27$ in Table~8 of \cite{2023MNRAS.518.1260S}, which corresponds to $\hat{\Delta}_{12k}$ in our notation.  We contaminate these selected inputs by adding  10 times larger  standard errors to the inputs. For example, the predicted time delay between lensed images $A$ and $B$ of quasar J0029-3814 with $68\%$ credible interval  in Table~8 of \cite{2023MNRAS.518.1260S} is $-100^{ +4.7}_{\small -3.0}$, which is modified to be $\hat{\Delta}_{AB1}+10\hat{\sigma}_{\Delta_{AB1}}=-100 + 47=-53$ and the $1\sigma$ uncertainty of this modified time delay input remains the same as 4.7.}

\textcolor{black}{The model fits are visualized in Figure~\ref{ex2res}. Overall, the meta-analyses with Student's $t$ error  robustly infer $H_0$ in all cases. The posterior distribution of $H_0$ under Gaussian error is not sensitive to the manipulation of 9 inputs (10\%), either. However, it begins to deviate from the true value with wider spread when 18 inputs (20\%) are contaminated. This might be because  the increased input data size (90 input pairs in this study compared to 40 in the previous one) makes the meta-analysis under Gaussian error less sensitive to the small number of manipulated data. When 27 inputs (30\%) are modified, five Markov chains under Gaussian error  start exploring local modes without jumping between modes. Thus,  we replace the Metropolis update for $H_0$ with the repelling-attracting Metropolis update to encourage the Markov chains to jump between modes. The white histogram in the bottom-right panel of Figure~\ref{ex2res} shows the multimodal posterior distribution  explored by the the repelling-attracting Metropolis within Gibbs sampler.  The true value of $H_0$ is located near a valley between the two highest modes, meaning that the meta-analysis under Gaussian error loses its ability to capture the true value of $H_0$ near the highest mode.} 

\begin{table}[b!]
\caption{Numerical summaries  obtained by fitting the meta-analysis with either Student's $t$ error or Gaussian error. \textcolor{black}{We report} the posterior means (Post.~mean), posterior standard deviations (sd),  bias in percentage, coefficient of variation (CV) in percentage,  and root mean square error (RMSE). The true value of $H_0$ is 70.}
\begin{center}
{\footnotesize\begin{tabular}{cccccc}
\hline
& &No outliers & 9 outliers (10\%)   & 18 outliers (20\%) & 27 outliers (30\%)\\ 
\hline

\multirow{2}{*}{Post.~mean (sd)}& Gaussian & 69.884 (0.513) &  69.823 (0.543)& 70.216 (0.665) & 69.609 (1.509)\\ 
&Student's $t$ & 69.897 (0.487)  &  69.921 (0.471)& 69.887  (0.507 )& 69.796  (0.521)\\ 
&&&&&\\

\multirow{2}{*}{Bias (\%)} &Gaussian & 0.166 &   0.253 & 0.308 & 0.559\\ 
 & Student's $t$ & 0.148 &  0.112 & 0.161 & 0.292\\ 

&&&&&\\
\multirow{2}{*}{CV (\%)} & Gaussian & 0.733 &  0.776 & 0.949 & 2.156\\ 
&Student's $t$&  0.695  &  0.672 & 0.724& 0.744\\
&&&&&\\
\multirow{2}{*}{RMSE} & Gaussian & 0.526 &  0.571 & 0.699 & 1.559\\ 
&Student's $t$&  0.497  &  0.477 & 0.519& 0.559\\
\hline
\end{tabular}}\label{table3}
\end{center}
\end{table}


\textcolor{black}{Table~\ref{table3} numerically summarizes the  model fits. Estimates and evaluation criteria obtained by both types of error do not show notable differences even when there are 18 outlying inputs, which is consistent to the first three panels in Figure~\ref{ex2res}. In the case where we manipulate 27 input time delays (30\%), bias under Gaussian error does not reflect the multimodal nature of the posterior distribution well because the posterior mean (69.609) is still near the true value (70).  However, both coefficient of variation  and root mean square error become about  three times larger under Gaussian error due to  the multimodal aspect of the resulting posterior distribution of $H_0$.}

\textcolor{black}{When it comes to the computational cost, it takes  about 2,900 seconds on average to \textcolor{black}{implement 10,000 iterations by the default Metropolis-Hastings within Gibbs sampler, while it takes about 3,400 seconds on average to implement repelling-attracting Metropolis within Gibbs sampler.  The average Gelman-Rubin diagnostic statistic for $H_0$, except the multimodal case of the Gaussian error with 27 outliers, is 1.011 and the largest value for this diagnostic statistic is 1.027. In the multimodal case,  the Gelam-Rubin diagnostic statistic with the default Metropolis-Hastings within Gibbs sampler is 3.629, while it reduces to 1.146 with the repelling-attracting Metropolis within Gibbs sampler. The  effective sample size for $H_0$ under Student's
$t$ error increases from 262 (out of 25,000) to 490  as the number of outliers increases, while it   decreases from 251 to 9 (23 when multimodal sampler is adopted) under Gaussian error. The posterior predictive checks do not show evidence for the lack of the model fit even for the multimodal case under Gaussian error because the posterior distribution in this case  is nearly centered at the true value; see  \textcolor{black}{Figure~2 of the second supplementary material} for more details.}
}



\section{\textcolor{black}{\textcolor{black}{Application to Realistic Data with  Limitations}}}\label{sec5.3real}

\textcolor{black}{We apply the proposed  meta-analysis to a realistic input data set composed of  three strong lens systems, 2M1134-2103, PSJ 1606-2333, and SDSS 1206+4332.  The first two lens systems are quadruply-lensed quasars, meaning that we see four lensed images of each quasar in the sky.  Their Fermat potential difference estimates are obtained from Table~8 of \cite{2023MNRAS.518.1260S} and time delay estimates are reported in \cite{millon2020six}. These two lenses have not been used to infer $H_0$ \textcolor{black}{because their  redshifts of the lenses ($z_d$) are not measured. Instead, the work of \cite{2023MNRAS.518.1260S}  fixes these redshifts  at their fiducial estimate, 0.5, and we also follow this approach. This is a limitation of the input data for the meta-analysis, and thus we  implement a sensitivity analysis. The result (though not reported here) shows that incorrectly setting $z_d=0.1$ for PSJ 1606-2333 instead of 0.5, which makes 3 input pairs out of 7 (43\%) wrong, does not affect the resulting posterior distribution of $H_0$ much; the shape is almost identical, and the posterior mean and standard deviation change by less than 5\% and 4\%, respectively.} The last system is a doubly-lensed quasar. We obtain its Fermat potential difference estimate from \cite{birrer2019sdss} \textcolor{black}{and  time delay estimate from \cite{2023ApJ...950...37M}. In the work of \cite{birrer2019sdss},} the Fermat potential difference estimate is not  reported as a numeric value, but its posterior distribution is visualized. Thus, we  set the estimate as the mode of the posterior distribution of $\phi_{AB}$, and $1\sigma$ uncertainty as the difference between the range of the posterior distribution divided by 4. These values are tabulated in \textcolor{black}{Table~2 of the first supplementary material.}}

\begin{figure}[b!]
\begin{center}
\includegraphics[scale = 0.145]{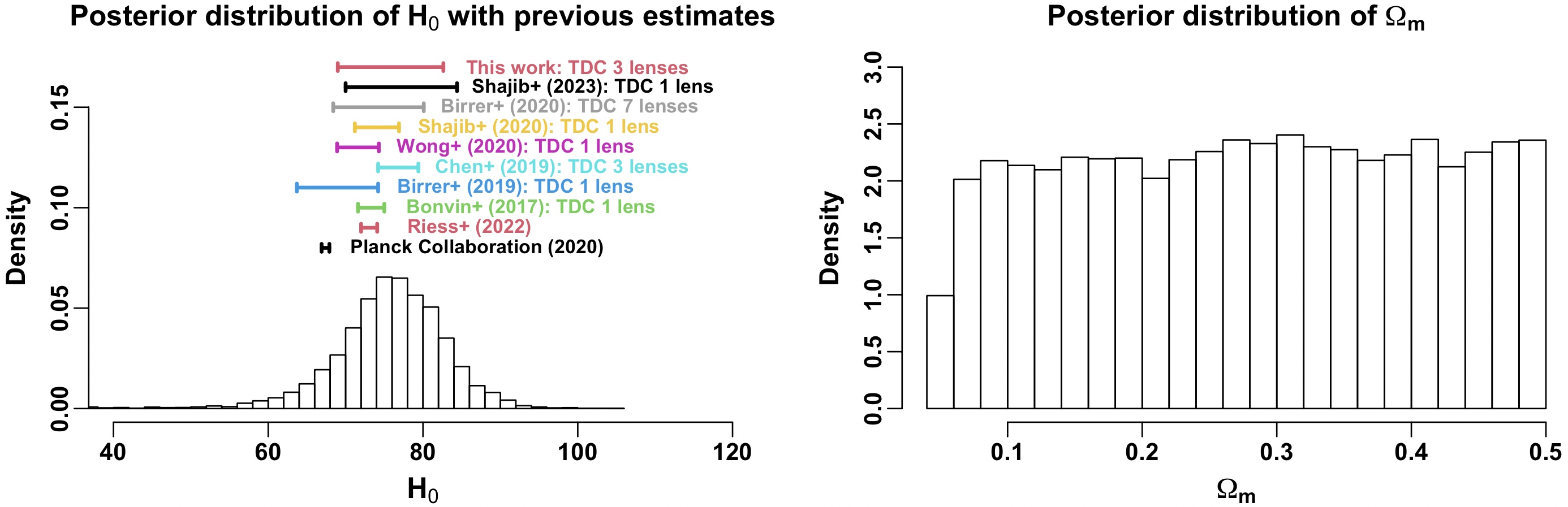}
\caption{\textcolor{black}{The left panel displays} the  marginal posterior distributions of $H_0$ with the three strong lens systems, 2M1134-2103, PSJ 1606-2333, and SDSS 1206+4332. The posterior mean and standard deviation are 75.81 and 6.82, respectively. Its coefficient of variation in percentage is 9\%. \textcolor{black}{We denote 68\% intervals from existing works above the posterior distribution of $H_0$. The spread of the posterior distribution of $H_0$ is wide enough to encompass all of these 68\% intervals, while its central location is  consistent to most of the estimates from time delay cosmography. The posterior distribution of $\Omega_{\text{m}}$ is almost uniformly distributed, which is related to the wide range of the posterior distribution of $H_0$ due to the small data size.}}
\label{ex3res}
\end{center}
\end{figure}

\textcolor{black}{Figure~\ref{ex3res} shows the  posterior distribution of $H_0$ \textcolor{black}{in the left panel  and that of $\Omega_{\textrm{m}}$ in the right panel}. The posterior mean and standard deviation \textcolor{black}{of $H_0$} are 75.81 and 6.82, respectively, which is 9\% coefficient of variation in percentage. \textcolor{black}{We note that this $H_0$ estimate involves two caveats; the deflector redshifts of the two quad-lenses in the meta-analysis are  set to their fiducial estimate, 0.5; and the meta-analysis does not model line-of-sight velocity dispersion.} For a comparison with existing estimates in tension, we display 68\% intervals on top of the distribution. The first eight intervals are based on the time delay cosmography \citep{bonvin2016he, birrer2019sdss, 2019MNRAS.490.1743C, wong2020, shajib2020, birrer2020tdcosmo, 2023A&A...673A...9S}, including this work. We denote how many lenses are used in each work.  The bottom two intervals represent  the $H_0$ estimates in tension between  the early and late Universe measurements, that is, $67.4 \pm 0.5$ \citep{planck2020cosmo} and $73.04 \pm 1.04$    \citep{2022ApJ...934L...7R}, respectively. It turns out that the spread of the posterior distribution from this work is wide enough to encompass all of the 68\% intervals, while its central location is  consistent to most of the estimates from time delay cosmography and is close to  the estimate from the late Universe measurement, $73.04 \pm 1.04$    \citep{2022ApJ...934L...7R}.}

\textcolor{black}{It is worth noting that the work of \cite{birrer2019sdss}, whose 68\% interval is $(63.7,~ 74.2)$ denoted in blue, uses only one doubly-lensed quasar,  SDSS 1206+4332. This quasar is also used in this work, but the middle value of the resulting 68\% interval of this work shown at the top of Figure~\ref{ex3res} is larger than that of \cite{birrer2019sdss}. This may be because the information about $H_0$ contained in SDSS 1206+4332  is averaged with the information in two other lens systems of this work. The interval of \cite{birrer2020tdcosmo}, colored in gray, also confirms this averaging effect because the impact of SDSS 1206+4332 is reduced when it is averaged with six other lenses, producing the larger central location. }

\textcolor{black}{The posterior distribution of $\Omega_{\text{m}}$ is almost uniformly distributed over the range of the uniform prior distribution, although it has slightly more mass near 0.3, 0.4, and 0.5, and less mass on small values near 0.05. This approximately uniform posterior distribution of $\Omega_{\text{m}}$ is partially related to the wide range of the posterior distribution of $H_0$, both of which  can be ascribed to the small sample size.}

\textcolor{black}{The computational cost is not expensive in analyzing these three lens systems.  It takes  about 48 seconds on average to implement 10,000 iterations.  The Gelman-Rubin diagnostic statistic for $H_0$ is \textcolor{black}{1.0034}. The  effective sample size for $H_0$ is \textcolor{black}{1,183} out of 25,000. From the posterior predictive check, we do not find particular evidence that the model is not sufficient to explain the data; see  \textcolor{black}{Figure~3 of the second supplementary material} for more details.}

\section{Concluding Remarks}\label{sec5}

Using time delay cosmography, we have proposed a \textcolor{black}{robust} meta-analysis \textcolor{black}{based on Student's $t$ error} to infer the current expansion rate of the Universe, called the Hubble constant ($H_0$).  Input data for this meta-analysis are the estimates of  time delays and Fermat potential differences that can be  obtained from independent studies in the literature. \textcolor{black}{Thus, the meta-analysis does not need to model the time series data or high-resolution imaging data from scratch to estimate time delays and Fermat potential difference estimates, respectively.} The output of this meta-analysis \textcolor{black}{is} posterior samples of two cosmological parameters, $H_0$ and  $\Omega_{\textrm{m}}$, \textcolor{black}{and external convergence of each lens system ($\kappa_{1}, \kappa_{2}, \ldots, \kappa_{K}$)}.  \textcolor{black}{Two} simulation \textcolor{black}{studies emphasize how robustly and accurately the proposed meta-analysis can infer $H_0$ in the presence of outliers.} In a realistic data analysis, we apply the meta-analysis to \textcolor{black}{three lens systems, two of which have never been used in estimating $H_0$ in the literature, and estimate $H_0$ by $75.81\pm6.82$. This corresponds to   9\% coefficient of variation ($100\hat{\sigma}_{{H}_0}/\hat{H}_0$), which is called precision level in time delay cosmography.} 

In the era of the Vera C.~Rubin Observatory Legacy Survey of Space and Time and James Webb Space Telescope,  an unprecedented number of strong lens quasars will be detected. For example, \cite{2010MNRAS.405.2579O} expect that the Rubin Observatory will detect thousands of lensed quasars. 
\textcolor{black}{Consequently,  more and more input pairs for the meta-analysis will become available via the development and improvement of automated data analytic tools.  The ability of the meta-analysis to collect and extract the  information about $H_0$, commonly embedded in these input estimates, will be of great importance as it can produce a robust benchmark estimate  of $H_0$ without requiring the analyses of time series and imaging data from scratch.}


\textcolor{black}{It is worth noting}  that the meta-analysis takes the Fermat potential difference estimates  as inputs, while most published articles that infer $H_0$ via time delay cosmography do not report   Fermat potential difference estimates.  \textcolor{black}{Even though Fermat potential values can be easily obtained once a lens model is fit,  most people tend to consider Fermat potential as an intermediate product, focusing more on the lens model  itself (e.g., Einstein radius, slope profile, etc.) or $H_0$ (or time delay distance $D_{\Delta}(H_0, z_k, \Omega)$). Consequently, Fermat potential difference estimates are frequently overlooked.}  \textcolor{black}{We hope that this work  encourages the community to recognize their importance and make it a standard practice to report these values.}
\begin{acks}[Acknowledgments]
HT appreciates Michael Fleck, a multimedia specialist in the department of statistics at  Pennsylvania State University, for creating the image in the left panel of Figure~\ref{fig1lensing}.  XD acknowledges the support from the  World Premier International Research Center Initiative (WPI) in Japan.   We thank Simon Birrer, \textcolor{black}{Tommaso Treu, the editor of physical science, associate editor, and  referees of the Annals of Applied Statistics} for their productive and insightful comments on this manuscript.
\end{acks}




\begin{supplement}
\stitle{The simulated and realistic data sets}
\sdescription{Tables of the simulation study and realistic data used in numerical studies.}
\end{supplement}
\begin{supplement}
\stitle{Posterior Predictive Checks}\label{supp2}
\sdescription{Posterior predictive checks for all of the numerical studies.}
\end{supplement}
\begin{supplement}
\stitle{R Code}
\sdescription{The R code, standalone.R, reproduces all of the results reported in this work. The same results can be obtained
by using the R package, \texttt{h0}.}
\end{supplement}





\bibliography{bibliography}
\bibliographystyle{imsart-nameyear}

\end{document}